\documentclass[aip,cha,amsmath,amssymb,reprint]{revtex4-1}

\usepackage{graphicx}
\usepackage{dcolumn}
\usepackage{bm}

\usepackage{xcolor}
\newcommand{\todo}[1]{}
\renewcommand{\todo}[1]{{\color{red}[[TODO: {#1}]]}}

\newcommand\lett[1]{(\textbf{#1})}
\newcommand{\setB}[0]{\ensuremath{\mathcal{B}}}

\newcommand{\ego}[0]{\ensuremath{\mathrm{ego}}}
\newcommand{\alter}[0]{\ensuremath{\mathrm{alter}}}

\newcommand{\rmA}[0]{\ensuremath{\mathrm{A}}}
\newcommand{\rmB}[0]{\ensuremath{\mathrm{B}}}
\newcommand{\rmC}[0]{\ensuremath{\mathrm{C}}}

\newcommand{\hcross}[2]{\ensuremath{h_{\times}(\mathrm{#1} \mid \mathrm{#2})}}

\newcommand{\lambQuot}[0]{\ensuremath{\Lambda_\mathrm{quote}}}
\newcommand{\lambRand}[0]{\ensuremath{\Lambda_\mathrm{random}}}

\begin{document}

\title[The quoter model]{The quoter model: a paradigmatic model of the social flow of written information}

\author{James P.~Bagrow}
\email{james.bagrow@uvm.edu}
\affiliation{Department of Mathematics \& Statistics,
University of Vermont, Burlington, VT 05405 USA
}

\author{Lewis Mitchell}%
 \email{lewis.mitchell@adelaide.edu.au}
\affiliation{School of Mathematical Sciences, North Terrace Campus, The University of Adelaide, SA 5005 Australia
}

\date[]{Received 31 October 2017; accepted 23 February 2018; published online 11 July 2018}

\begin{abstract}
We propose a model for the social flow of information in the form of text data, which simulates the posting and sharing of short social media posts.
Nodes in a graph representing a social network take turns generating words, leading to a symbolic time series associated with each node.
Information propagates over the graph via a quoting mechanism, where nodes randomly copy short segments of text from each other.
We characterize information flows from these text via information-theoretic estimators, and we derive analytic relationships between model parameters and the values of these estimators.
We explore and validate the model with simulations on small network motifs and larger random graphs.
Tractable models such as ours that generate symbolic data while controlling the information flow allow us to test and compare measures of information flow applicable to real social media data.
In particular, by choosing different network structures, we can develop test scenarios to determine whether or not measures of information flow can distinguish between true and spurious interactions, and how topological network properties relate to information flow.
\end{abstract}


\maketitle

\begin{quotation}
Rich datasets on human activity and behavior are now available thanks to the widespread adoption of online platforms such as social media.
The primary artifact generated by users of these platforms is text in the form of written communication.
These symbolic data are invaluable for research on information flow between individuals and across large-scale social networks, but working with and modeling natural language data is challenging.
While most models of social information flow focus on compartment models, contagion models, or cascades, the richness of the text data available to researchers underscores the importance of incorporating the full information present in text into modeling efforts.
In this paper we propose a model for how groups of individuals embedded in a social network can generate streams of text data based on their own interests and the interests of their neighbors in the network.
The goal is to more explicitly capture the dynamics inherent to human discourse.
We show how to relate parameters in the model to quantities underlying information-theoretic estimators specifically aimed at understanding information flow between sources of text.
By controlling the graph topology and model parameters we can benchmark how information flow measures applied to text deal with spurious interactions and confounds.
\end{quotation}

Recently, considerable effort has taken place to better understand information flow
in dynamical systems and real datasets~\cite{e16095068}.
On one hand, new measures and algorithms have been developed to better understand information flow interactions and related phenomena, including transfer entropy~\cite{schreiber2000measuring}, symbolic transfer entropy~\cite{staniek2008symbolic}, convergent cross-mapping~\cite{sugihara2012detecting}, and causation entropy~\cite{sun2014causation,sun2015causal}.
On the other hand, new large-scale datasets have allowed researchers to better understand at scale the spread of information in a complex system, especially those involving online social networks and social media such as Twitter~\cite{gruhl2004information,iribarren2009impact}.
Especially interesting are studies applying information-theoretic tools to large-scale social media data, such as Ver Steeg and Galstyan, who consider the shared information present in the timings of tweets posted by social ties on Twitter~\cite{ver2012information}, and Borge-Holthoefer \emph{et al.}, who use symbolic transfer entropy to investigate predictive signals of collective action such as protests in the time series of the numbers of tweets posted in different geographic regions~\cite{borge2016dynamics}.
These recent studies show that tools developed from information theory and dynamical systems theory can successfully be applied to human dynamics data captured from online platforms such as Twitter.

Most research on information flow within online media either considers proxies of information flow, such as tracking the spread of particular keywords, or uses information-theoretic tools focused on the timing of social media posts~\cite{ver2012information,borge2016dynamics}. 
Yet the posts themselves are packed with potentially useful data: the text generated by users of online platforms is their primary artifact and, when available for study, should be the focus of research. Fortunately for the study of information flow, information theory has a rich history of working with symbolic data such as text.

Given the importance of focusing on the text data, there is currently a lack of models for the problem of studying information flow as measured from the text generated by users in a social network. 
Most work focuses on modeling information flow as a type of contagion, cascade or diffusion process~\cite{burt1987social,watts2002simple,gruhl2004information,vespignani2012modelling}. 
These works are invaluable for studying information flow but by compartmentalizing nodes into groups that have or have not adopted an innovation, been ``infected'', etc. they generally neglect the full richness of the text generated by users in this setting.

Our goal here is to propose and analyze a simple model of the discourse underlying the text generation process online. 
Nodes within a given graph (representing individuals within a social network) generate symbolic time series (the time-ordered text) based on what they and their neighbors in the network say, and we relate this to information-theoretic estimators of information flow between the texts of different individuals.
Doing so provides insights into how well these estimators can distinguish true versus spurious interactions, detect confounding effects, and help us relate network topological properties to the features of information flow.

The rest of this paper is organized as follows.
In Sec.~\ref{sec:background} we discuss background material on entropy estimators for written text and how they may be used to measure information flow.
In Sec.~\ref{sec:themodel} we introduce the quoter model and discuss its different components.
In Sec.~\ref{sec:ecaanalysis} we analyze the quoter model between two individuals and compare our analytic predictions with simulations.
Sec.~\ref{sec:quoteronnetworks} extends these simulations to a number of network structures and investigates the interplay between network topology and information flow.
We conclude with a discussion of our results and potential future directions in Sec.~\ref{sec:discussion}.

\section{Background}
\label{sec:background}

\subsection{Entropy and information flow in text}

The information content in a written text can be quantified with its entropy rate $h$, the number of additional bits (or other unit of information) needed on average to determine the next word~\footnote{In this work we consider the word-level entropy rate, but it is also common to work on a per-character basis.} of the text given past words~\cite{CoverThomas}. 
The entropy rate is maximized for a text that is completely random such that preceding words will not give useful information for determining a subsequent word. 
Conversely, the entropy rate is zero for a deterministic sequence of words such that knowledge of previous words only gives all the information necessary to specify the subsequent word.

There is a rich history of practical entropy estimators for text~\cite{shannon1951prediction,ebeling1994entropy,schurmann1996entropy}.
The challenge when working with real text is that there is information in the ordering of words, not just their relative frequencies---shuffling a text preserves the (unigram) Shannon entropy but destroys much of the information in the text. 
To account for the ordering of words, one needs to evaluate the complete joint (or conditional) distribution of word occurrences, and estimating these probabilities requires enormous amounts of data.

Kontoyianni et al.~\cite{kontoyiannis_nonparametric_1998} proved that the estimator
\begin{equation}
    \hat{h} = \frac{T \log_2 T}{\sum_{t=1}^{T} \Lambda_{t}} = \frac{\log_2 T}{\bar{\Lambda}},
    \label{eqn:hhat}
\end{equation}
converges to the true entropy rate $h$ of a text,
where $T$ is the length of the sequence of words and $\Lambda_{t}$ is the \emph{match length} of the prefix at position $t$: it is the length of the shortest substring (of words) starting at $t$ that has not previously appeared in the text. 
(For simplicity, we now omit the $\hat{}$ symbol distinguishing the estimator from the true quantity.)
Theorems underlying nonparametric estimators such as Eq.~\eqref{eqn:hhat} play an important role in the mathematics of data compression. 
Indeed, some authors have even used compression software to estimate the entropy of text.
However, using compression software risks introducing bias, as specific compression code (such as gzip) trades off optimal compression rates in order to run much more efficiently.
Due to these trade offs, one should instead work directly with the theoretical estimator (Eq.~\eqref{eqn:hhat}) to more accurately estimate $h$.

Equation~\eqref{eqn:hhat} generalizes naturally to a \textbf{cross-entropy} between two sequences $A$ and $B$~\cite{243444,bagrow2017information}.
To do so, define the \emph{cross-parsed match length} $\Lambda_{t}(A | B)$ as the length of the shortest substring starting at position $t$ of sequence $A$ not previously seen in sequence $B$. 
If sequences $A$ and $B$ are \emph{time-aligned}, as in a written conversation unfolding over time, then `previously' refers to all the words of $B$ written prior to the time when the $t$th word of $A$ was written. 
The estimator for the cross-entropy rate is then
\begin{equation}
	h_{\times}(A \mid B) = \frac{T_{A} \log_2 T_{B}}{\sum_{i=1}^{T_{A}} \Lambda_{i}(A \mid B)},
	\label{eqn:crossEntropy}
\end{equation}
where $T_{A}$ and $T_{B}$ are the lengths of $A$ and $B$, respectively. 
The log term in Eq.~\eqref{eqn:crossEntropy} has changed to $\log_2 T_{B}$ because now $B$ is the ``database'' we are searching over to compute the match lengths and the $T_{A}$ factor is due to the average of the $\Lambda_{t}$'s taking place over $A$. 
The cross-entropy tells us how many bits on average we need to encode the next word of $A$ given the information previously seen in $B$.
Further, $h_{\times}(A \mid A) = h$.
Despite a similarity in notation, the cross-entropy is distinct from the conditional entropy (which requires estimating a joint probability distribution of $A$ and $B$, something that is not easy to estimate from social media text data, for example).
The cross-entropy can be applied directly to text of a pair of individuals by choosing $B$ to be the text stream of one individual and $A$ the text stream of the other.
 
While our focus in this work is on the cross-entropy between pairs of individuals, $h_\times$ can be generalized further to $\hcross{A}{\setB}$, quantifying the predictive information regarding the text in string $A$ contained within a \emph{set} of strings $\setB$~\cite{bagrow2017information}.
This lets us understand the information flow from multiple social ties to a single individual.
It also allows us to construct transfer entropy-\emph{like} measures: $h(A) - \hcross{A}{\{A,B\}}$  measures how much 
if any extra information is present on average in the past text of $B$ about the future future text of $A$, beyond
the information already present in the past text of $A$.
Doing so is important when inferring information flow from data, as it is important to determine whether or not the information in $B$ is redundant if one already has the information in $A$~\cite{schreiber2000measuring,sun2014causation,sun2015causal}.

\subsection{Social information flow}

In a previous work we showed how to use the cross-entropy (Eq.~\eqref{eqn:crossEntropy}) as a measure of information flow between individuals posting to the Twitter.com social media platform~\cite{bagrow2017information}.
We concatenated the texts of all public tweets for a given Twitter user into a long stream of text, and then applied
the aforementioned entropy and cross-entropy measures to users, pairs of users, and ego-centric networks consisting of users and their most frequent contacts.
Measuring information flow with the cross-entropy naturally incorporates the temporal ordering of the tweet text and uses all the available information in the texts of the individuals, whereas other measurement methods limit themselves to proxies of information flow, such as tracking the spread of keywords like hashtags or URLs.

The focus of that work was on measuring information flow from text data.
When developing and applying estimators in such scenarios it is useful to have plausible models with which to build examples and test cases. 
However, most work modeling information flow has focused on the study of information as ``packets'' spreading between individuals, typically represented in Twitter's case by the hashtags or URLs. 
This allows researchers to apply contagion models, such as Susceptible-Infected or other compartmental models, complex contagion models, and more~\cite{burt1987social,weng2012competition,toole2012modeling,melnik2013multi}. 
Contagion models are very well studied on network topologies, but in this case they neglect the dynamical processes governing written communication. 
The back-and-forth nature of discussions, for example, may generate far more information flow within the text than would be measurable from the spread of keywords alone.

\section{The quoter model}
\label{sec:themodel}

We propose the ``quoter model'' as a simplified way to capture the dynamics governing the written conversations taking place between individuals in a social network.
The model consists of $N$ individuals embedded as the nodes $\mathcal{V}$ of a social network $\mathcal{G} = (\mathcal{V},\mathcal{E})$ where $|\mathcal{V}| = N$ and there are $|\mathcal{E}| = M$ edges connecting those nodes.
For generality we take the graph to be directed such that an edge $(i,j) \in \mathcal{E}$ represents communication from node $j$ to node $i$ via the quoting process described below.

\begin{figure}
\centering
\includegraphics[width=\columnwidth]{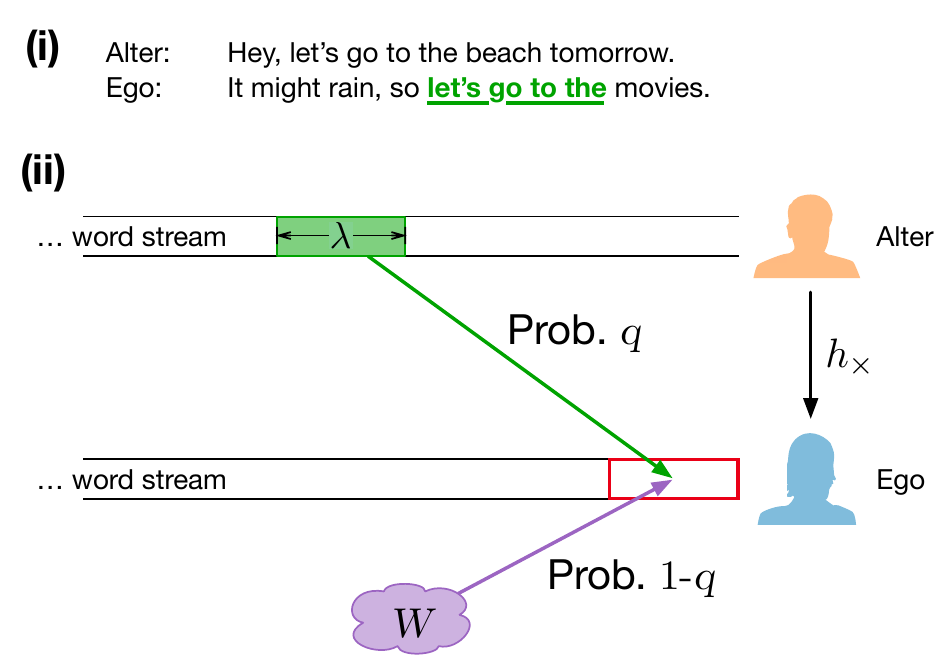}
\caption{The quoter model for the social flow of information.
\lett{i} The repeated occurrences of short quoted passages such as this one throughout a written conversation indicate information flow.
\lett{ii} In the model, words are generated by individuals at each time step, forming word streams. To model information flow we use two mechanisms: at each timestep, with probability $1-q$ the ego draws $\lambda$ new words randomly from a specified vocabulary distribution $W$; otherwise, with probability $q$ the ego copies a passage of length $\lambda$ taken from a random position in the past words of the alter. 
\label{fig:cartoon}}
\end{figure}

Each member of the graph generates written text over time, represented as a symbolic time series or ``word stream''. 
At timestep $t$, individual $i$ generates a number of new words according to one of two mechanisms, growing his or her word stream.
The number of new words at timestep $t$ is $\lambda_i(t) \sim L_i(t)$, where this number is drawn from an integer-valued length distribution $L_i(t)$. 
This probability distribution may be time-independent or evolve as a function of time, and this distribution may vary across users ($L_i \neq L_j, j \neq i$) or not ($L_i = L_j \equiv L$).
After choosing the number of words to generate, the actual words are generated according to one of two mechanisms:
\begin{enumerate}
\item $\lambda_i(t)$ draws with replacement from a vocabulary distribution $W_i$ (with probability $1-q_{ij}$);
\item a contiguous sequence of $\lambda_i(t)$ words are copied from a random position within the previously written text of a neighbor $j$ of node $i$ (with probability $q_{ij}$).
\end{enumerate}
This process is then repeated for all individuals in the network until their text streams have reached a desired length or a desired number of timesteps have elapsed.
The first mechanism is intended to represent the creation of new content while the second mechanism is the quoter action of the model. 
The quote probabilities $q_{ij}$ tune the relative strengths of the two mechanisms by how often $i$ quotes from the past text of $j$.
We illustrate one step of the model for a pair of individuals in Fig.~\ref{fig:cartoon}.

The idea underlying the second mechanism is that when two individuals are discussing a topic verbally or in writing, and they are listening to one another, then there will be a back and forth of small sequences of common words. The quotes generated by the second mechanism are not meant to capture full-length, long form quotations such as retweets, but instead short shared sequences of text. Alice: ``That's the right way to go''; Bob: ``No, this is the right way.''. In this example, the exchange between Alice and Bob leads to a short quotation of Alice by Bob (``the right way'') and from this exchange only we can at least surmise that Bob is probably receiving and reacting to Alice's text.
Of course, Bob could have responded in an equivalent way without that short quote. However, over the course of very long conversations we expect more such quotations to occur on average, and they will likely occur more often in conversations when there is more information flow than in conversations where there is little information flow.

\subsection{Model components}
\label{subsec:modelcomponents}

The main components of the quoter model are (i) the graph topology, which may be as simple as a single directed link between two individuals, (ii) the quote probabilities $q_{ij}$, (iii) the length distributions $L_i$, and (iv) the vocabulary distributions $W_i$.
We study several graph topologies in this work.
The quote probabilities $q_{ij}$ can be considered as edge weights on the social network, and there is considerable flexibility in assigning those weights.

The length distributions $L_i$ govern the amount of text generated per timestep and the total length of the text: the expected length after $t$ timesteps will be $\left<L\right> \times t$. We primarily focus on two cases here, the constant length distribution $L(\lambda_t) = \delta_{\lambda\lambda_t}$, where $\delta_{ij}$ is the Kronecker delta; and a Poisson distribution $L(\lambda_t) = e^{-\lambda}\lambda^{\lambda_t}/\lambda_t!$ with mean $\lambda$.

The vocabulary distribution $W_i$ gives the relative frequencies of words for individual $i$. In this work we consider two example $W$'s.
The first is a uniform distribution over a fixed number of $z$ unique words: $W(w) = 1/z, w = 1, \ldots, z$. The binary case corresponds to $z=2$. 
The second vocabulary distribution  is a basic Zipf's law that incorporates the skewed distributions typically observed in real text corpora~\cite{zipf1949human}.
Here the probability of a word $w$ depends on its rank $r_w$, with the most probable word having rank $r_w=1$. 
Zipf's law then defines word probabilities that obey a power law form with $r$:
$W(w) \propto r_w^{-\alpha}$, where $\alpha$ is a power law exponent.
This distribution is normalized by $H_{z,\alpha} = \sum_{r=1}^{z} r^{-\alpha}$, the generalized harmonic number.
This distribution also holds for infinite vocabularies ($z \to \infty$) so long as $\alpha > 1$, in which case the normalization constant converges to the Riemann zeta $\zeta(\alpha)$.

\section{Model analysis}
\label{sec:ecaanalysis}

Here we study the basic quoter model between two individuals (referred to as the ``ego'' and the ``alter'') where the ego copies the alter but the alter does not copy the ego. 
We focus on the case of uniform vocabulary distribution $W(w) = 1/z, w=1,\ldots,z$ and
we assume both individuals draw from the same $W$, although our analysis is not specific to these assumptions.

To quantify the flow information from the alter to the ego via the cross-entropy $\hcross{ego}{alter}$, we need to compute the mean ${\Lambda} = T^{-1} \sum_{t=1}^{T} \Lambda_t$, where $\Lambda_t$ is the length of the shortest substring of words beginning at position $t$ in the ego's text which has not been observed in the text of the alter prior to ``time'' $t$ (Sec.~\ref{sec:background}), and $T$ is the total length of the text.
To model $\Lambda_t$, we assume that (i) two terms contribute to $\Lambda_t$: the mean $\Lambda$ when a quote occurs (call it $\lambQuot$) and the mean $\Lambda$ when no quote occurs (call it $\lambRand$); and (ii) the quote probability $q$ weights these two possibilities: 
\begin{equation}
{\Lambda_t}(\ego \mid \alter) = (1-q)\lambRand + q \lambQuot,
\label{eqn:lambdabaroverall}
\end{equation}
where we have suppressed the dependence on position $t$ in $\lambRand$ and $\lambQuot$.
We need to determine both $\lambRand$ and $\lambQuot$ as functions of the vocabulary distribution and the current amounts of text generated.

\subsection{Prefix matches when not quoting}

It is possible as the ego is drawing words from the vocabulary distribution that due to chance a string of words will be generated that previously appeared in the past text of the alter. This will depend on the vocabulary distribution and the length of the alter's past text.

Suppose the alter has posted a total of $t$ words so far and the ego has just posted $m$ new words. 
The probability that one of the new words posted by the ego matches a random word previously posted by the alter is 
$\sum_w W(w)^{2} \equiv d$.
This is the probability that two draws from the vocabulary distribution give the same word, irrespective of the particular word, and is the Simpson index (also known as the Herfindahl-Hirschman index) of the vocabulary distribution\footnote{
The vocabulary distribution $W(w)$ enters into our analysis through its diversity $d$, the probability of drawing the same word twice irrespective of the particular word $w$. 
In the case of text obeying Zipf's law, the diversity is $d = \zeta\left(2\alpha\right)/\zeta\left(\alpha\right)^{2}$, where $\zeta$ is the Riemann zeta function.
For $\alpha=3/2$, which gives a reasonable approximation for the vocabulary distribution of real text~\cite{williams2015text} although there is evidence that Zipf's law is too simplistic for real text corpora~\cite{williams2015zipf,williams2015text}, we find that $d\approx 0.176$. 
Note that when Zipf's law is defined over a finite vocabulary of $z$ words, the diversity becomes $d = H_{z,2\alpha} / \left(H_{z,\alpha}\right)^{2}$.
Again for the plausible exponent $\alpha = 3/2$, and a vocabulary size of $z=1000$ total words, this gives $d \approx 0.185$, slightly higher than the infinite case.
These two examples show that real vocabulary can tend to relatively high values of $d$ despite the large number of words, because the probabilities for those words are so heavily skewed.
}$^{,}$\footnote{The diversity can also account for individuals with different vocabulary distributions.
In that case, $d = \sum_w W(w)^{2}$ becomes $d = \sum_w W_i(w) W_j(w)$. 
This only requires that both distributions are defined over the same set of words, which can always be achieved simply by taking any words missing in $W_i$ ($W_j$) but present in $W_j$ ($W_i$) and including them in $W_i$ ($W_j$) with a probability of zero.}.
The probability of at least $m$ new ego words matching with $m$ prior alter words at a particular location in the alter's past text is $d^{m}$. 
Since there are approximately $t$ locations in the alter's text at which a match may occur (assuming $t \gg m$), the expected number of matches of length $m$ or more is $t d^{m} \equiv C(m)$. 
Then the expected length of the longest match $m_*$ occurs at the value of $m=m_*$ for which $C(m_*)\geq 1$ and $C(m_*+1) < 1$.
Solving $C(m_*) = 1$ for $m_*$ gives an expected longest match length of 
$m_* = \ln(t) / \ln(1/d)$, 
or
\begin{equation}
\lambRand = \frac{\ln(t)}{\ln(1/d)} + 1
\label{eqn:lambRand}
\end{equation}
since $\Lambda$ is always one more than the match length.

\subsection{Prefix matches when quoting}

If a quote of length $\lambda$ occurs at position $t$, then $\Lambda_t = \lambda + 1$ only if any  words of the ego subsequent to the $\lambda$ quoted words do not happen to match the words of the alter subsequent to the original quoted passage.
In other words, even if deterministically a match of length $\lambda$ occurs, $\Lambda_t$ may be longer due to chance.
Specifically, the probability that $\Lambda_t = \lambda + 1 + m$, $m\geq0$, is $d^{m}(1-d)$,
as a value of $m$ requires that the next $m$ words will match and the $(m+1)$-th word will not match.
Note that, unlike the previous calculations, this probability does not involve the total text length of the alter $t$ because these post-quote matches cannot occur anywhere in the alter's text except in the positions following the quoted passage (neglecting duplicate passages).
From this probability, the expected $\Lambda_t$ is
\begin{equation}
\sum_{m=0}^{\infty}  \left(\lambda+1+m\right) d^{m}(1-d)= \lambda + 1 + \frac{d}{1-d},
\end{equation}
meaning that, on average, random chance increases $\Lambda_t$ by an amount $\frac{d}{1-d}$.

However, it is not necessarily reasonable to neglect duplicate passages.
Indeed, the number of duplicate passages may be significant for certain combinations of parameters: the probability that a different location of the alter's past is the start of a passage of length $\lambda$ equal to the randomly chosen quoted passage is $d^{\lambda}$, and the expected number of such duplicate passages within the alter's text (including the original passage) is $\approx t d^{\lambda} + 1$. 
For $t=10^{4}$, $d=1/5$, and $\lambda=3$, for example, the expected number of duplicates is 17.

The probability for at least $m$ words of the ego's text subsequent to the newly quoted passage to also match $m$ words following the original passage in the alter is $d^{m}$, so the expected number of times matches of length $m$ or longer will occur following any of the duplicate passages in the alter is $\approx (t d^{\lambda}+1)d^{m}$.
The longest match length $m_\ast$ occurs at the value of $m$ for which the number of these matches is ${\approx}1$, or $m_\ast = \ln(t d^{\lambda}+1)/\ln(1/d)$.
Lastly, the expected total match length when quoting is then $\lambda + \ln(t d^{\lambda}+1)/\ln(1/d)$. 

\begin{figure}[t!]
\centering
\includegraphics[width=\columnwidth]{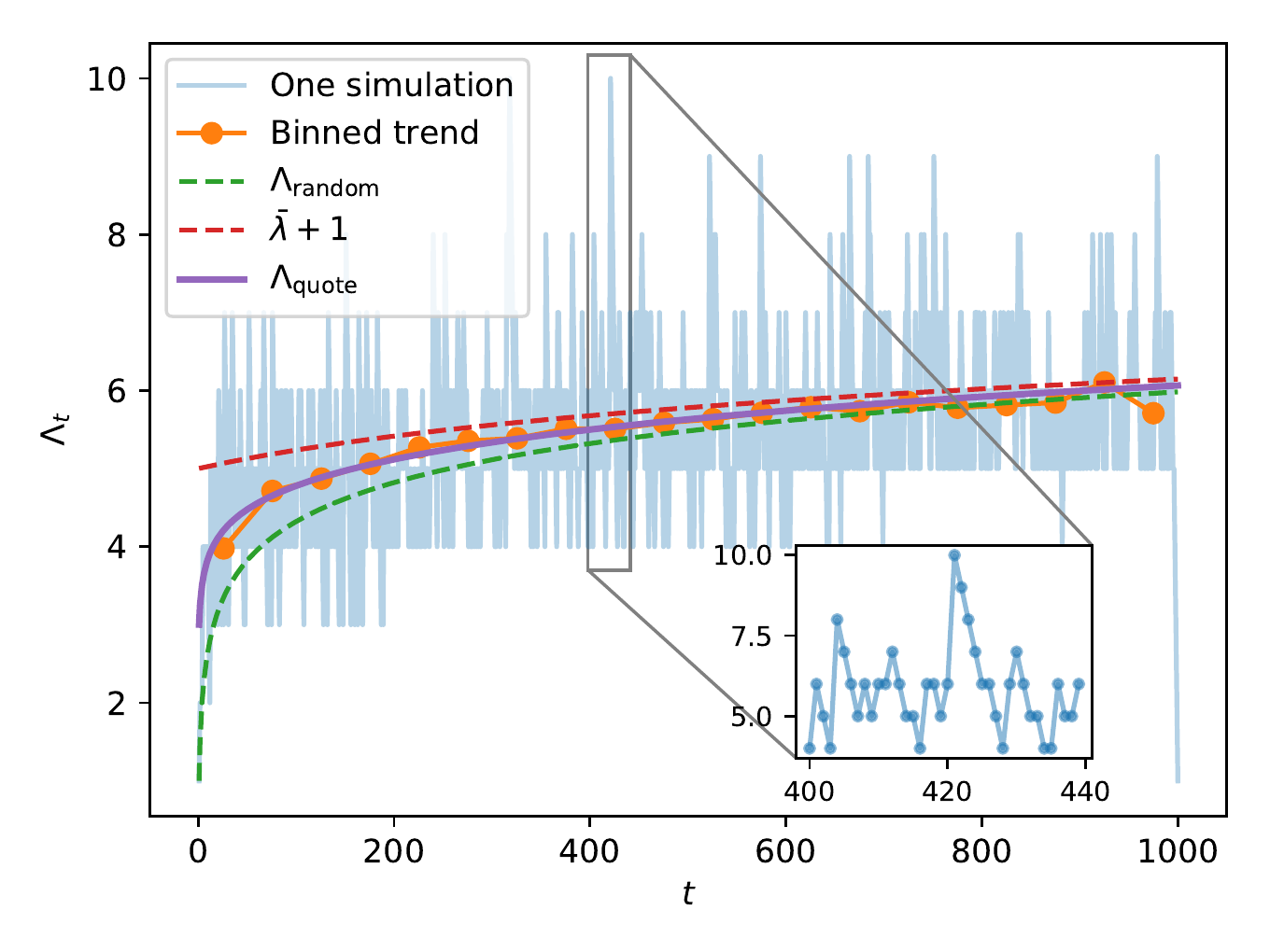}
\caption{Illustration of $\Lambda_t$ when quoting to demonstrate the relationship between $\lambRand$, $\bar{\lambda}$, and $\lambQuot$. 
Here we show a single realization of the model
(parameters: $q=1, z = 4, \lambda=4$). 
Individual realizations show considerable variability so we also include a binned trend averaged over $n=10$ realizations.
This trend agrees well with $\lambQuot$ for these parameters.
The inset plot highlights a spike in $\Lambda_t$ (at $t = 421$) and how it decays linearly back to the approximate level of $\lambRand$.
\label{fig:simLamb}}
\end{figure}

However, unlike with $\lambRand$, adding 1 to this expected total match length is not an accurate expression for the average $\lambQuot$.
When $\lambda + \ln(t d^{\lambda}+1)/\ln(1/d)$ is much larger than $\lambRand$, then the match length $\Lambda_t$ at that text position $t$ will almost certainly be due only to the single quoted passage.
This means that the subsequent $\Lambda_{t+1}$ will likely be 1 fewer than $\Lambda_t$, because a random match that would extend $\Lambda_{t+1}$ is unlikely. Likewise, $\Lambda_{t+2} = \Lambda_t-2$, and so forth, until the match lengths are short enough that random matching is again probable.
Accounting for this, we expect the average $\lambQuot$ to be roughly equal to
\begin{equation}
\frac{1}{\bar{\lambda}-\lambRand+2} \sum_{j=0}^{\bar{\lambda}+1-\lambRand} \left(\bar{\lambda} + 1 - j\right),
\end{equation}
where $\bar{\lambda} = \lambda + \frac{\ln(t d^{\lambda}+1)}{\ln(1/d)}$.
Equivalently, this is the average of the two endpoints, $\bar{\lambda}+1$ and $\lambRand$, and therefore:
\begin{equation}
\lambQuot = \frac{1}{2}\left(\lambda + \frac{\ln(t d^{\lambda}+1)}{\ln(1/d)} + \frac{\ln(t)}{\ln(1/d)}+2\right).
\end{equation}
We illustrate the relationship between $\lambRand$, $\bar{\lambda}$, and $\lambQuot$ in Fig.~\ref{fig:simLamb}, showing a single simulation of the model and highlighting a spike in $\Lambda_t$ above $\lambRand$ and how it decays back down to $\lambRand$.

With these expressions for $\lambRand$ and $\lambQuot$ we can now compute $\Lambda$ and from it the cross-entropy.

\subsection{Cross-entropy}

To compute the cross-entropy $h_\times$ between the ego and alter requires computing the total $\Lambda$ summed over all positions in the ego's text where matches can occur then dividing $T \ln T$ by that $\Lambda$: $h_\times = T \ln T / \Lambda$, where $\Lambda = \sum_{t=1}^{T} \Lambda_t$.
Using the previously derived expected contributions to $\Lambda$ for the two mechanisms and approximating the sum over the text positions with an integral gives the following  expression for $\Lambda$:
\begin{align}
\Lambda \approx& \int_{0}^{T}\Big[ \left(1-q\right)\lambRand + q \lambQuot \Big] \,  dt \nonumber \\
\begin{split}
     =&  \frac{T}{\ln(1/d)} \left[ (1-q) \left(\ln\frac{T}{d}-1\right)  \right. \\
       & + \left.   \frac{q}{2} \left(\ln \frac{T}{d^{\lambda+2}} + \left(\frac{1}{T d^{\lambda}}+1\right)\ln(T d^{\lambda}+1)-2 \right)\right],
     \label{eqn:integralForLambda}
     \end{split}
\end{align}
which can be substituted into $T \ln T / \Lambda$ to compute the cross-entropy as a function of $q$, $\lambda$, $d$, and $T$.

The limit of large text using Eq.~\eqref{eqn:integralForLambda} gives 
\begin{align}
\lim_{T \to \infty} \hcross{ego}{alter} &= \lim_{T\to\infty} \frac{T \ln T}{ \Lambda}
= \ln(1/d),
\label{eqn:limitInfiniteText}
\end{align}
which is the R\'enyi entropy of the vocabulary distribution:
\begin{equation}
h_\alpha = \frac{1}{1-\alpha}\ln\left(\sum_{w} W(w)^\alpha\right)
\end{equation}
with $\alpha = 2$.
Note also that $q$ has dropped out of this limit, implying that, given sufficient text, the entropy of the model will be that of the underlying vocabulary distribution only.
However, as we shall see, for finite $T$, even quite large, $q$ still plays an important role in the overall cross-entropy.

\subsection{Comparison with simulations}

\begin{figure}
\centering
{\includegraphics[width=\columnwidth]{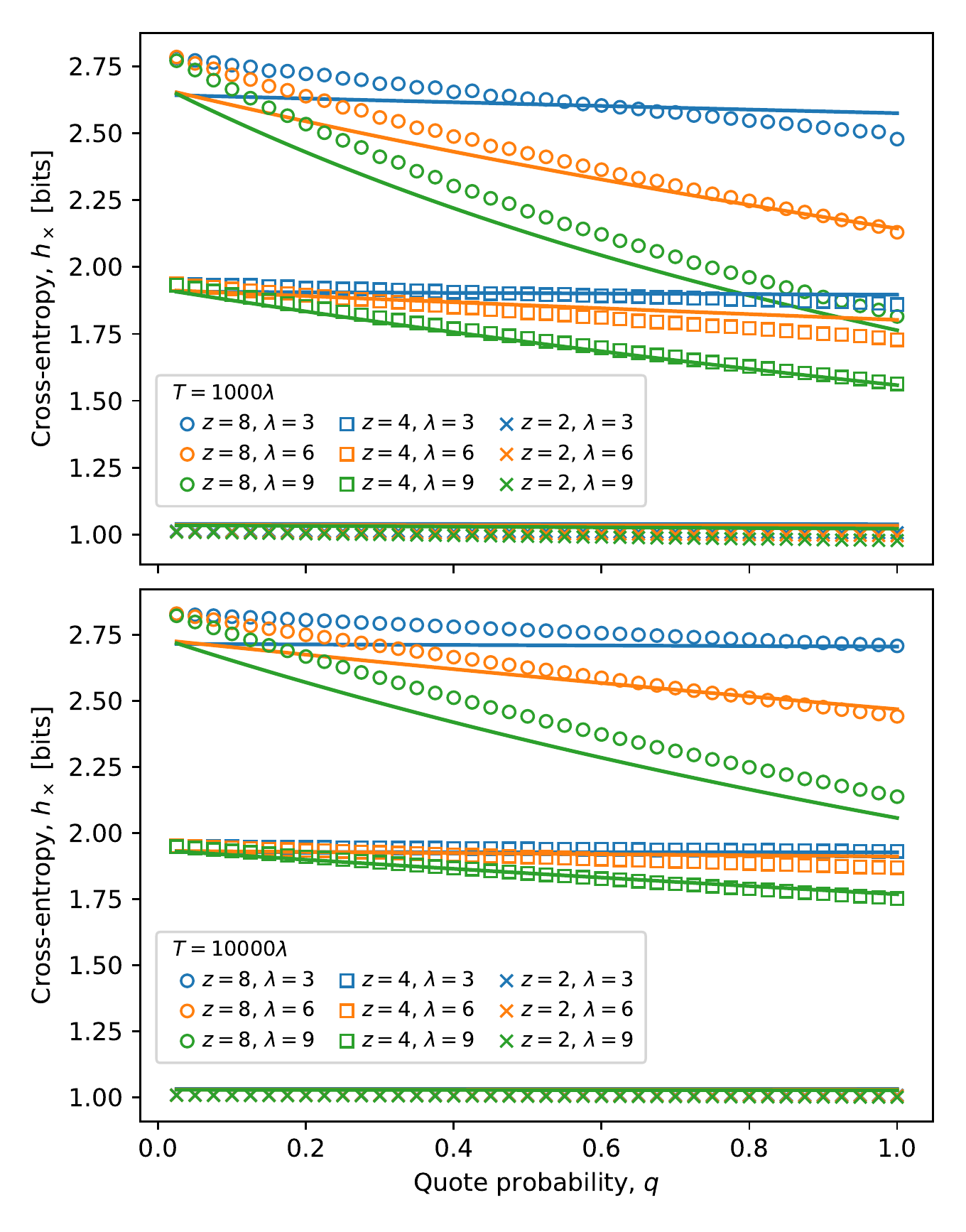}}
\caption{The theoretical predictions (lines) give qualitative agreement with simulations (symbols), although there are systematic discrepancies, especially at lower vocabulary diversities $d=1/z$.
\label{fig:validateH-vs-q}
}
\end{figure}

To test our theoretical predictions, we simulate the quoter model and compared our predicted cross-entropy (substituting Eq.~\eqref{eqn:integralForLambda} into $T \ln T / \Lambda$ and converting to bits) with that computed directly from the simulations (Eq.~\eqref{eqn:crossEntropy} on the simulated text sequences). 
We simulate the one-link, two-node model for $10^{3}$ and $10^{4}$ timesteps, giving expected text lengths of $T = 10^{3}\lambda$ and $T = 10^{4}\lambda$, respectively. 
Here we choose for both nodes $W(w) = 1/z$, $w=1,\ldots,z$, $L(t) = \mathrm{Pois}(\lambda)$, $q_{ij}=q$ and $q_{ji}=0$ (denoting the ego as $i$ and the alter as $j$).
Overall, we find reasonable qualitative agreement between our predictions and the simulations, as shown in Fig.~\ref{fig:validateH-vs-q}. However, there are some systematic discrepancies. 
While the absolute difference in entropies between predictions and simulations is small, often less than 0.1-0.2 bits, this means that the treatment above does not capture everything present in the model.

\begin{figure}
\centering
{\includegraphics[width=\columnwidth]{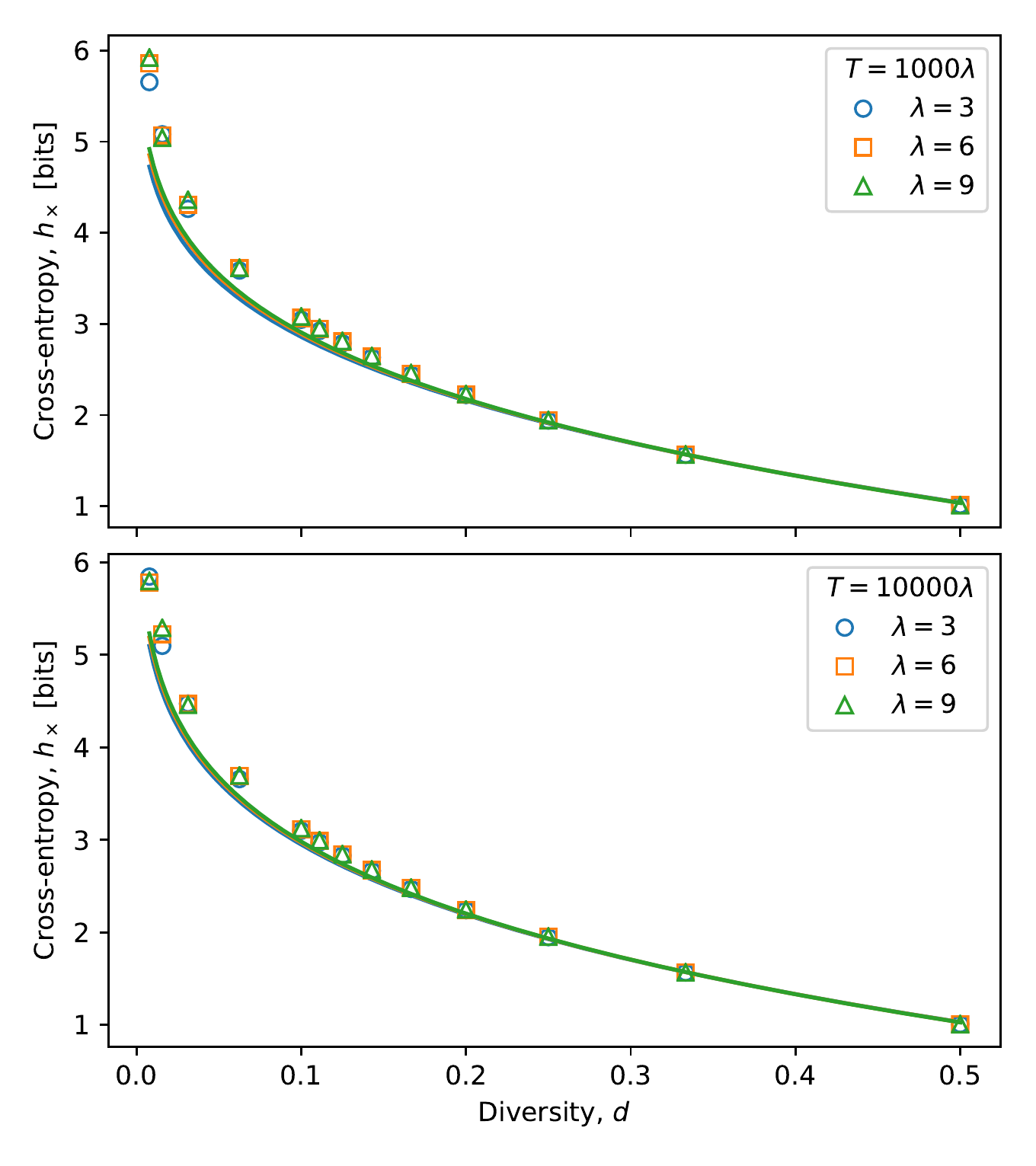}}
\caption{The limiting case of $q=0$ for different levels of vocabulary diversity $d = \sum_w W(w)^{2}$.
There is reasonable agreement between cross-entropy using Eq.~\eqref{eqn:integralForLambda} and simulations except at low values of $d$. Agreement improves with larger $T$.
\label{fig:validateH-vs-d_q0}
}
\end{figure}

Beyond Fig.~\ref{fig:validateH-vs-q}, which explores the cross-entropy as a function of $q$ for different $\lambda$ and $d=1/z$ parameters, 
it is also useful to inspect the two limiting cases of no quotes ($q=0$) and all quotes ($q=1$).
Figure \ref{fig:validateH-vs-d_q0} explores how the cross-entropy depends on $d$ when $q=0$. 
Since there are no quotes, we expect no dependence on $\lambda$ and we indeed see strong collapse across the simulations and the theory (there is a slight difference between the curves only because the total length of the generated text depends on $\lambda$).
Further, there is good agreement with predictions (solid lines) except at values of very low $d$ (equivalently, high $z$). Agreement improves considerably at higher $T$ although predicted values are still below those of the simulations.
In this case, $h_\times$ depends entirely on $\lambRand$, and the expression for $\lambRand$ (Eq.\eqref{eqn:lambRand}) primarily gives only the \emph{scaling} of $\lambRand$ with accuracy.

\begin{figure}
\centering
{\includegraphics[width=\columnwidth]{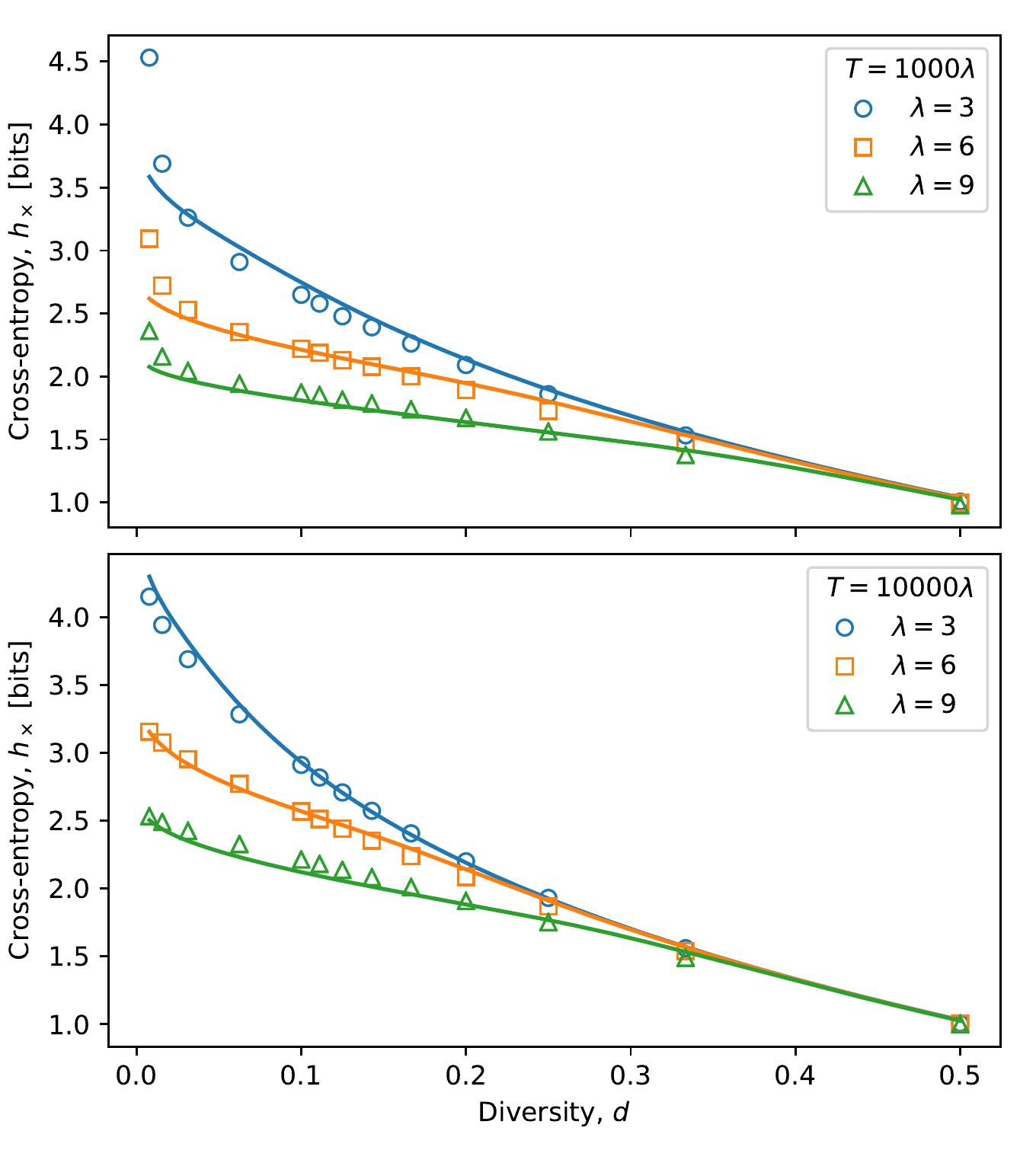}}
\caption{The limiting case of $q=1$ for different levels of vocabulary diversity $d$. Symbols denote simulations and lines denote predicted cross-entropy using Eq.~\eqref{eqn:integralForLambda}.
Agreement is reasonable in this case, and agreement improves for larger $T$.
\label{fig:validateH-vs-d_q1}
}
\end{figure}

The all-quote case is explored in Fig.~\ref{fig:validateH-vs-d_q1}. 
In this case, we expect a strong dependence on $\lambda$ and indeed we see a change of more than two bits of cross-entropy at the lower diversity values when moving from $\lambda=3$ to $\lambda=9$.
We also see good agreement between predictions and simulations except at low $d$, although in this case agreement improves considerably at low $d$ for the longer text length.

\medskip

Overall, we find that our treatment of the model captures the basic qualitative links between $q$, $d$, $\lambda$ and the total text length.
Agreement is not perfect, indicating that more behavior is going on than is being modeled, particularly at low $d$, or entropy estimators based on $\Lambda$ are biased for finite text, or some combination thereof.
A more rigorous treatment of the model may be able to distinguish between these two possibilities, and can extend the analysis to more complex arrangements than a single link between a pair of individuals.

\section{The quoter model on networks}
\label{sec:quoteronnetworks} 

\begin{figure}
\centering
\includegraphics[width=0.9\columnwidth]{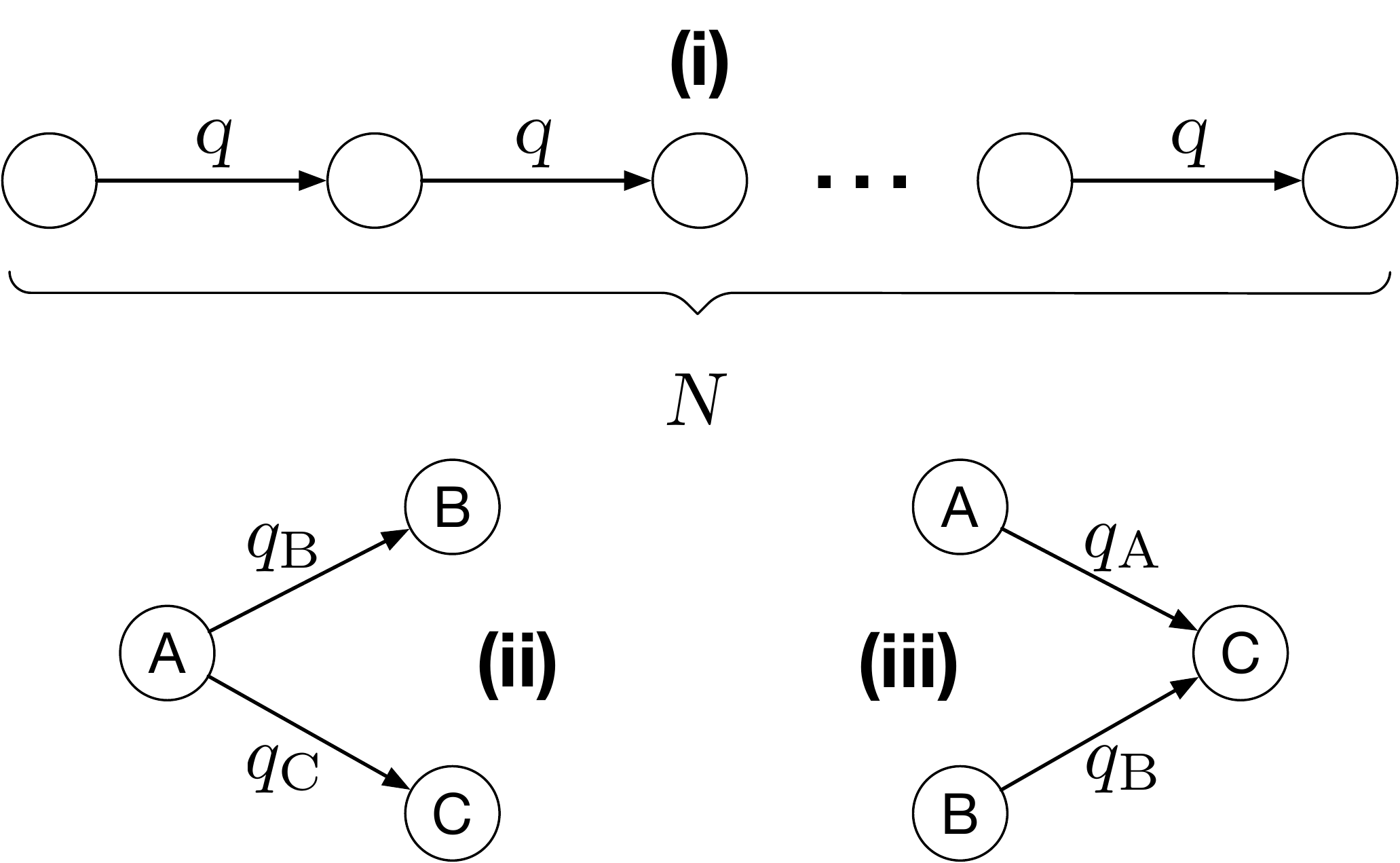}
\caption{Model network topologies. 
\lett{i} Chain of $N$ quoters, each with unidirectional quote probability $q$; 
\lett{ii} Fork, with quote probabilities $q_\rmB$ and $q_\rmC$ for nodes B and C to copy A respectively; 
\lett{iii} Collider, with quote probabilities $q_\rmA$ and $q_\rmB$ for C to copy A and B, respectively.
\label{fig:networks}}
\end{figure}

Moving beyond our treatment of a single pair of individuals (Sec.~\ref{sec:themodel}), 
here we numerically investigate the quoter model on four simple network topologies (see Fig.~\ref{fig:networks}):
A \emph{chain} of $N$ nodes where each node copies from the previous node (i),
a \emph{fork} where one node influences two nodes (ii),
a \emph{collider} where a node is influenced by two nodes simultaneously (iii),
and larger Erd{\H o}s-R\'enyi and Barab{\'{a}}si-Albert random graphs (iv) (not shown in Fig.~\ref{fig:networks}).
These topologies allow us to better understand, in a simplified context, the interplay between network topology and the dynamics of information flow as measured via the cross-entropy. 
The chain allows us to understand the attenuation of information flow with distance, the fork and the collider provide simple motifs to investigate confounds and spurious links, and the larger graph models can shed light on how global network properties such as density can affect information flow.

\subsection*{(i) Chain of quoters}

We investigate the attenuation of information by simulating the quoter model over a unidirectional chain of nodes $v_0, v_1,  \ldots, v_{N-1}$,
where each node has probability $q$ of quoting the node directly before them in the chain, except for the first node in the chain which only draws from $W$: 
\begin{equation}
q_{ij} = \begin{cases}
          q, & \text{if  $i>0, i=j+1$};  \\
          0, & \text{otherwise}.
\end{cases}
\end{equation}
At each timestep, each node in the chain writes or quotes $\lambda_t \sim \mathrm{Pois}(\lambda = 3)$ words, each of which is then
drawn from a 1000-word truncated Zipf distribution with exponent $\alpha = 3/2$.
(Results were found to be very similar when using a uniform distribution with the same number of words.)
We simulate the model on $N=10$ nodes for 10000 timesteps, so $T \approx 10000\lambda$.

\begin{figure}
\centering
\includegraphics[width=\columnwidth]{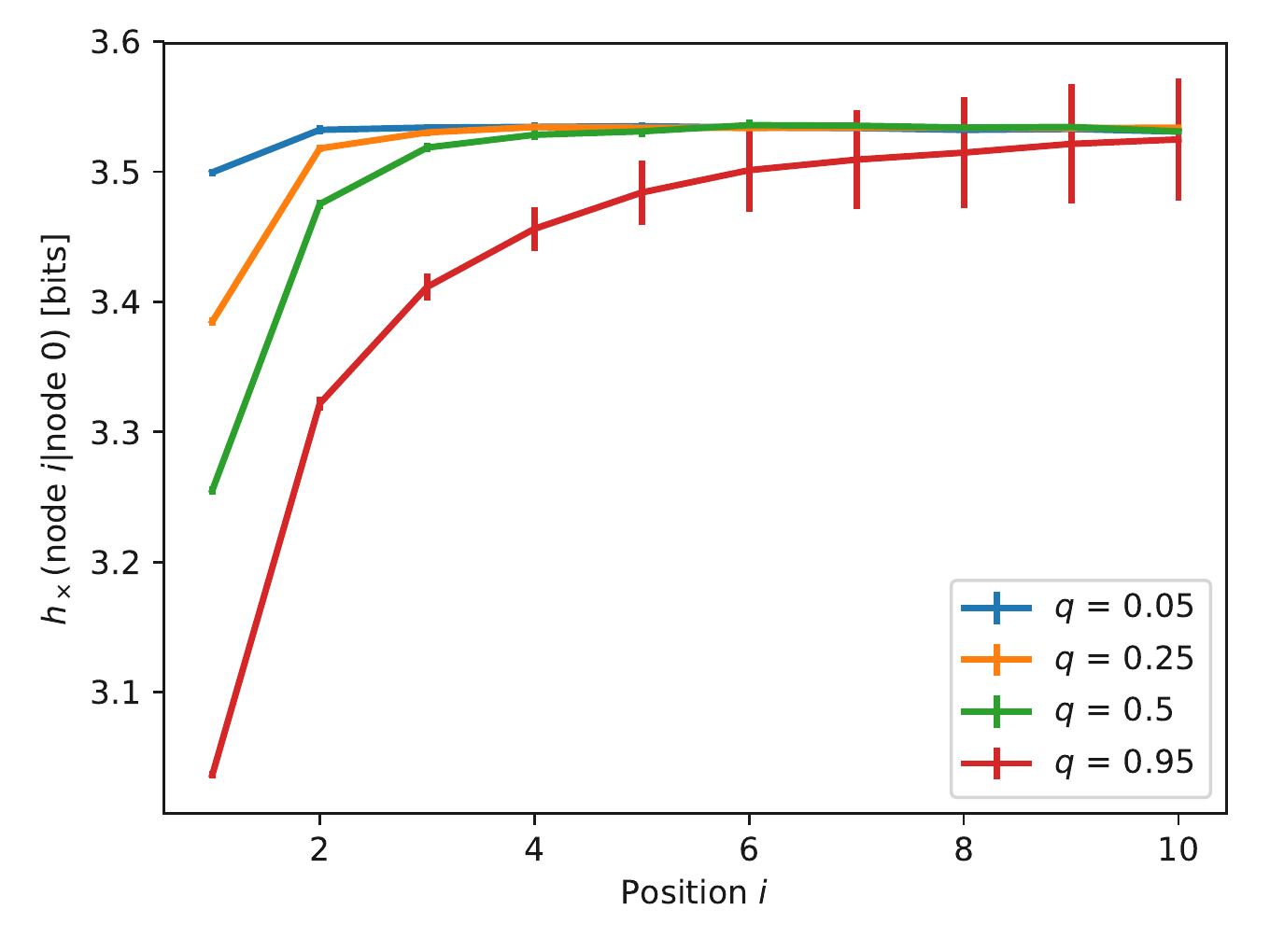}
\caption{Attenuation of information in a chain of quoters.
Cross-entropy increases (information flow decreases) with both distance from the source node and decreasing quote probability $q$, generally saturating for $q \leq 0.5$ by a separation of no more than four steps.
\label{fig:telephone}}
\end{figure}

Figure~\ref{fig:telephone} shows the cross-entropy of node $i$ from the first node 0 in the chain,
which generates original text.
For reasonable values of the quote probability $q < 0.5$ information attenuates quickly,
with $h_\times$ having saturated by approximately the third link in the chain.
Only at very high quoting probabilities ($q = 0.95$) do we observe greater information flow (lower cross-entropy) for nodes further along the chain.

\subsection*{(ii) Fork \& (iii) Collider}

\begin{figure}
\centering
\includegraphics[width=\columnwidth]{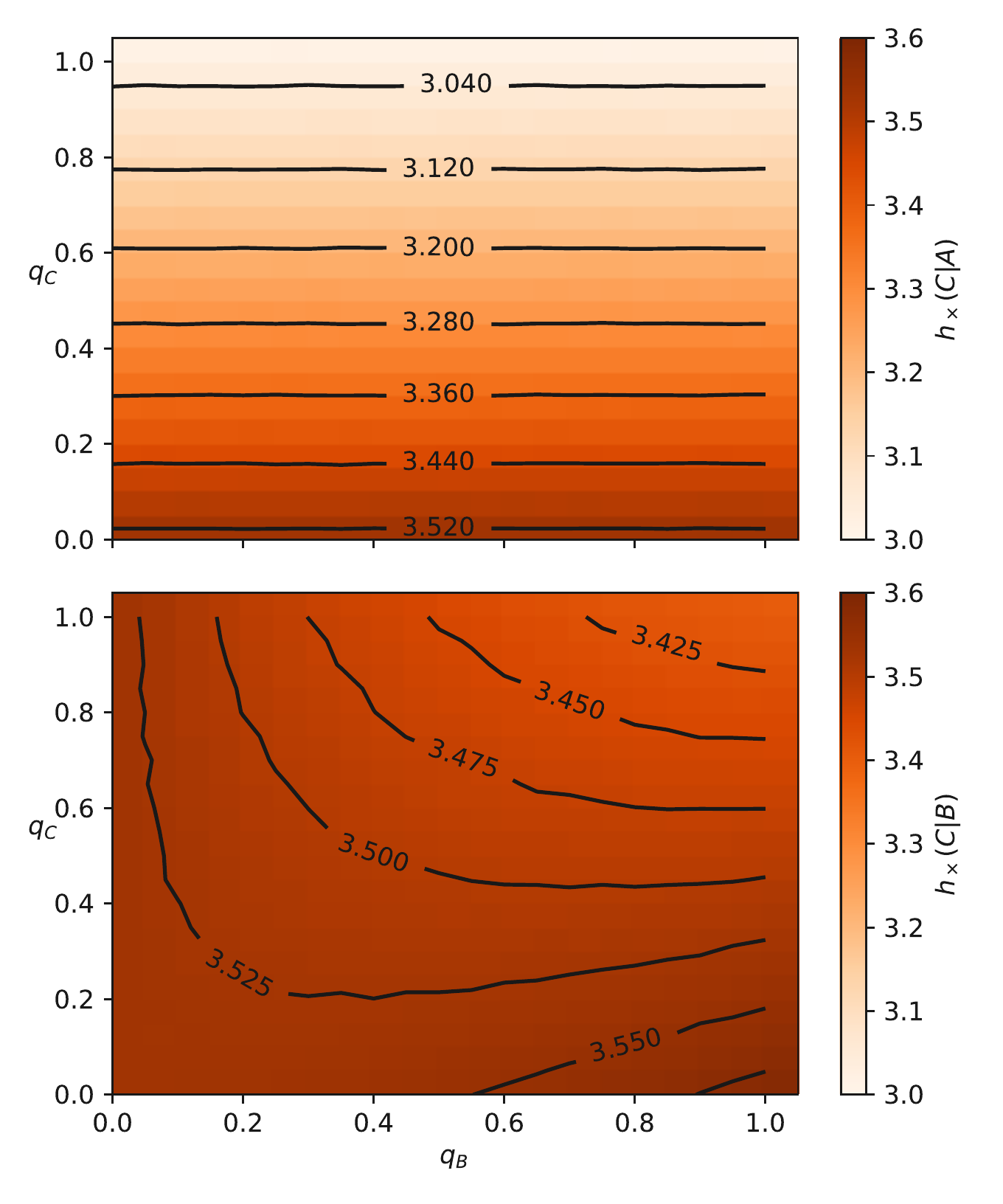}
\caption{Information flow on the fork graph as a function of the quote probabilities $q_\rmB$ and $q_\rmC$. 
(top) We find that the information flow between the source A to a target (either B or C; C shown only) depends only on the quote probability for the source-target link.
(bottom) The target-target link (B to C or C to B; B to C shown only) shows a mixed dependence on $q_\rmA$ and $q_\rmB$. However, the cross-entropy values are higher than those observed for the source-target links, for most regions of the $(q_\rmC,q_\rmB)$-space.
We discretized $q_\rmB$ and $q_\rmC$ into steps of 0.05 and interpolated to obtain the level curves in the figures.
\label{fig:fork}}
\end{figure}

To investigate how cross-entropy distinguishes between information flow from different sources, we simulate the quoter model on the three-node ``fork'' and ``collider'' graph shown in Fig.~\ref{fig:networks}.
First, for the fork graph (Fig.~\ref{fig:networks}(ii)), using the same parameters as above ($\lambda_t \sim \mathrm{Pois}(\lambda=3)$, $w \sim \mathrm{Zipf}(z = 1000, \alpha=3/2)$), we vary the probabilities $q_\rmB$ and $q_\rmC$ with which nodes B and C, respectively, copy the source node A, which generates original content (drawing words from $W$ only).
The top and bottom panels of Fig.~\ref{fig:fork} show the cross-entropy of C from A and of C from B, respectively, averaged over 1000 realizations of the model.
As expected, $h_\times(\rmC \mid \rmA)$ shows no dependence on $q_\rmB$ and decreases approximately linearly as the quote probability $q_\rmC$ grows (Fig.~\ref{fig:fork}; top).

The dependence of C upon B in the fork is more complex, however,
with the cross-entropy $h_\times(\rmC \mid \rmB)$ of the non-existent link between B \& C decreasing with both increasing $q_\rmB$ and $q_\rmC$ (Fig.~\ref{fig:fork}; bottom).
However, there exists a clear separation in the values of cross-entropy between the two cases,
with $h_\times(\rmC \mid \rmB)$ being significantly larger than $h_\times(\rmC \mid \rmA)$ for most quote probabilities except the region where both $q_\rmB$ and $q_\rmC$ are close to 1. 
Cross-entropy therefore effectively identifies the direction of real information flow for this model graph.

Due to the fork's symmetry, the results for $h_\times(\rmB \mid \rmA)$ and $h_\times(\rmB \mid \rmC)$ are identical to those shown in Fig.~\ref{fig:fork}.
Likewise, the analogous $h_\times(\rmC \mid \rmA)$ and $h_\times(\rmC \mid \rmB)$ for the collider network topology (Fig.~\ref{fig:networks}(iii)) appear similar to the top panel of Fig.~\ref{fig:fork}: with no dependence between A and B in the collider, $h_\times(\rmC \mid \rmA)$ decreases linearly with $q_\rmA$ and shows no dependence on $q_\rmB$ (not shown).

\subsection*{(iv) Random networks}

\begin{figure}
\centering
\includegraphics[width=\columnwidth]{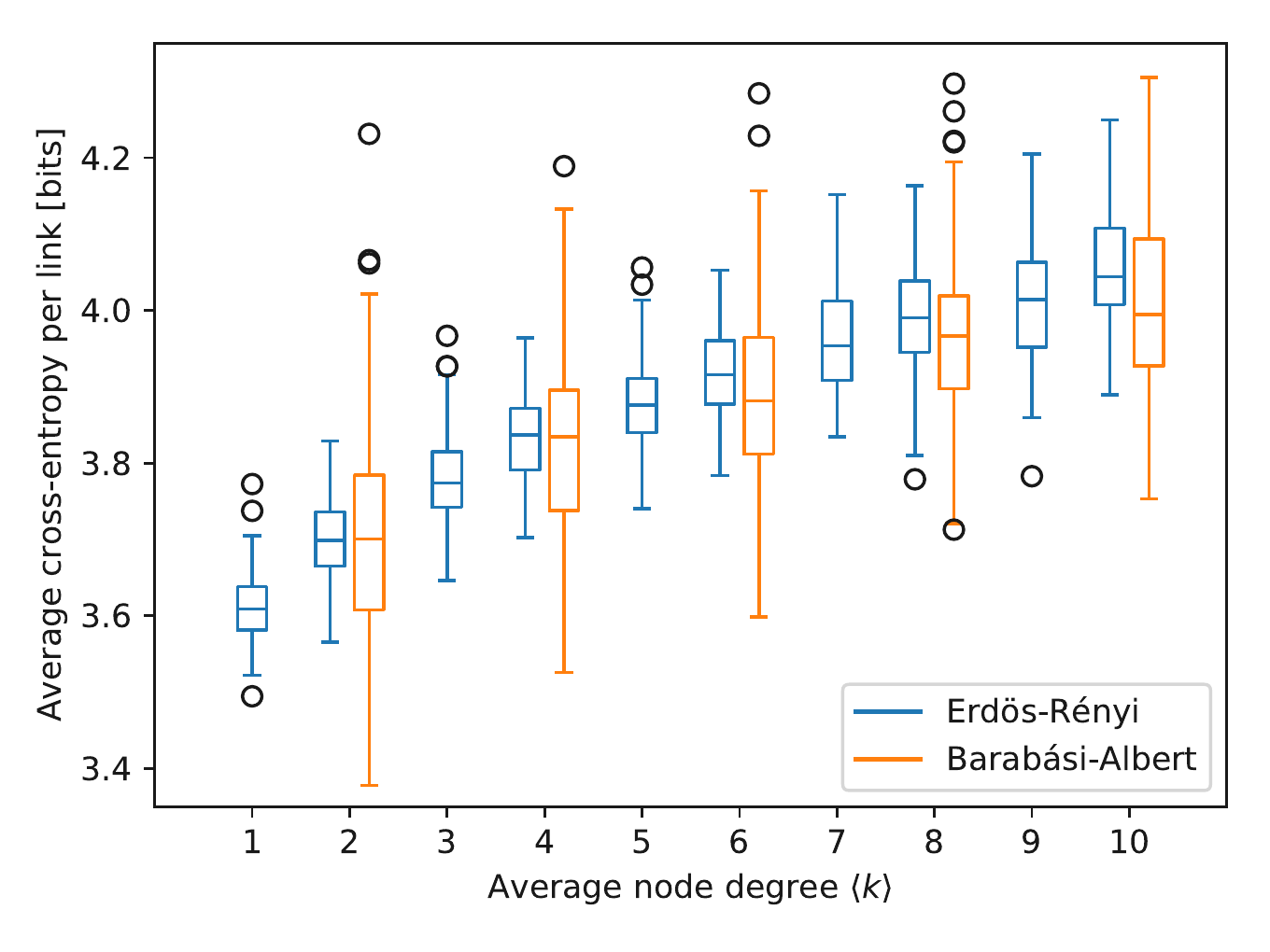}
\caption{Average information carried by edges in a network decreases as network density increases, as evidenced by the increase in cross-entropy. 
Both types of networks contain 100 nodes, and boxes represent the distribution of cross-entropy over 100 realizations each.
(Boxplots have been shifted left and right where they would otherwise overlap, for clarity.)
\label{fig:ERsim}}
\end{figure}

Finally, we investigate the quoter model on larger networks, modeled as random graphs.
We simulate the quoter model on Erd{\H o}s-R\'enyi (ER)~\cite{erdosrenyi1959,erdos1960evolution} and Barab{\'{a}}si-Albert (BA)~\cite{barabasialbert1999} random graphs.
ER graphs are simple models that capture only the overall density of a network, but are a useful starting point.
BA graphs capture the ``scale-free'' property observed in real-life social networks.
Using graphs of $N = 100$, we create directed, weighted networks of varying average node degree~\footnote{This was done for ER graphs by varying the probability $p$ for each link to exist. For BA graphs, we varied the attachment parameter $m$, which leads to only even-valued average degrees (see Fig.~\ref{fig:ERsim}).}.
To create directed ER networks, we chose pairs of nodes $i$ and $j$, and created an edge from $i$ to $j$ with probability $p$.
For the BA networks we used the standard preferential attachment method with edges pointing in both directions.
This construction means that quoting is always bidirectional in the BA networks, but not necessarily in the ER networks.
Other options are possible for the BA network, e.g., creating directed links from newer nodes to older nodes through the preferential attachment process, however this would have rendered these networks a directed tree rather than graph, as was desired here.

Quote probabilities $q_{ij}$ are chosen from $U(0,1)$,
with $q_{ii} \sim U(0,1)$ representing the probability of a node generating new content (after normalizing such that $q_{ii} + \sum_j q_{ij} A_{ij}$ = 1, where $A$ is the adjacency matrix of the graph). 
The quoter model is then run for $5000 N$ timesteps over the network, updating a randomly chosen node at each timestep, and using the same vocabulary ($W$) and quote-length ($L$) distributions as above.
At the end of the simulation each node has generated text of length $T \approx 5000 \lambda = 15000$ words.
We simulate 100 realizations of the network and quoter model dynamics on both the ER and BA networks.

Information flow on these graphs as a function of the graph's average node degree $\langle k \rangle$ is shown in Fig.~\ref{fig:ERsim}.
As average degree increases in the network, the average cross-entropy of a node $i$ from its neighbors $j$ also increases, meaning that $i$ becomes less predictable from its neighbors with increasing density.
The BA graphs show slightly lower median cross-entropy, however, with larger variation across realizations.
The presence of high-degree hubs in BA graphs means that cross-entropy can exhibit a larger range of variation, 
with the self probability $q_{ii}$ at hub nodes $i$ to generate new content driving much of the information flow on the network. 
The increasing trend of cross-entropy with average node degree indicates that information ``sources'' and ``sinks'' become increasingly difficult to identify in a network, 
as the density of connections increases.

\section{Discussion}
\label{sec:discussion}

In this paper, we introduced the quoter model as a simple, paradigmatic model of the flow of information.
Considerable effort has been put into measuring information flow in online social media, both from proxies such as tracking keywords and from information-theoretic tools. 
Models of the dynamics underlying these processes are invaluable for better understanding information flow, and the goal of our work is to introduce a model that more directly relates to information flow in text data than traditional contagion-style models, but without being overly complicated.
Our model mimics at a basic level the overall dynamics of text streams posted online, and here we showed that one can derive expressions for the information flow between written texts as measured via the cross-entropy.

The analysis we performed here showed good qualitative agreement with simulations in general, but there remains room for improvement. Nevertheless, the ability to find tractable expressions for information-theoretic quantities highlights how the basic quoter model can provide better insights into information flow over social networks. 
Indeed, we proposed this model because empirical benchmarks for information flow over social networks are difficult to find.
However, as many dynamic processes can be represented by symbolic time series, models like the quoter model may even be useful when studying information flow in more general contexts.

The language generator we studied here is a relatively simplistic bag-of-words model: individuals simply draw words from a given vocabulary distribution $W$.
More realistic models should be explored.
One possibility would be a time-dependent $W$.
For example, one could endow $W$ with a latent context $C$: $W(w \mid C)$ and allow the context to vary (slowly) over a space of contexts.
A Markov chain over this context space would be one way to introduce dynamic context shifts.
Such a context dependence can then be used to model topical shifts over the length of a discourse.
If two users exhibit the same context shifts, their vocabulary distributions will tend to ``sync up'' with each other, and this should lead to a lower cross-entropy than if contexts were not shared.

This dynamic context shift in quoted discourse suggests a natural time-based generalization to the model as well.
With quoting behavior likely to occur within a short ``attention span'' of the time of the original message,
it makes sense to incorporate a probability of quoting into the model which decays over time.
While the form of this probability likely introduces an extra parameter,
it is plausible that this parameter could be estimated from real data.
Future work will explore the possibility of fitting the quoter model to real datasets.

Lastly, there is much room for future exploration of network topology and its relationship to information flow.
As the quoter model allows us to design ``planted'' interactions, we can implement the quoter dynamics on constructed networks and then test whether algorithms can successfully infer true interactions and reject spurious interactions.
We did this here with the fork and collider graphs.
Moving beyond those small motifs, one area of network structure worth exploring in future work is that of network topologies exhibiting clustering, to investigate the effect of community structure~\cite{girvan2002community} on information flow.

\begin{acknowledgments}
We gratefully acknowledge the resources provided by the Vermont Advanced Computing Core at the University of Vermont and the Phoenix HPC service at the University of Adelaide.
This material is based upon work supported by the National Science Foundation under Grant No.\ IIS-1447634.
LM acknowledges support from the Data To Decisions Cooperative Research Centre (D2D CRC), and the ARC Centre of Excellence for Mathematical and Statistical Frontiers (ACEMS).
\end{acknowledgments}

\end{document}